\documentclass[aps,article,prl,preprintnumbers,amsmath,amssymb,twocolumn,superscriptaddress,showpacs,floatfix]{revtex4}

\usepackage{epsfig}
\usepackage{dcolumn}
\usepackage{bm}
\usepackage{color}

\usepackage{amssymb}
\usepackage[figuresright]{rotating}
\usepackage{wrapfig}

\newcommand{\be}{\begin{equation}}
\newcommand{\ee}{\end{equation}}
\newcommand{\bea}{\begin{eqnarray}}
\newcommand{\eea}{\end{eqnarray}}

\begin{document}

\title{Azimuthal Jet Tomography of Quark Gluon Plasmas at RHIC and LHC}
\author{Barbara Betz}
\affiliation{Institute for Theoretical Physics, Johann Wolfgang 
Goethe-University, 60438 Frankfurt am Main, Germany}
\author{Miklos Gyulassy}
\affiliation{Department of Physics, Columbia University, New York, 10027, USA}

\begin{abstract}

Recent data on the azimuthal and transverse momentum dependence  
of high-$p_T>10$~GeV pion nuclear modification factors in nuclear 
collisions at RHIC/BNL and LHC/CERN are analyzed in terms
of a wide class of jet-energy loss models and a variety of 
transverse expanding collective flow backgrounds. 
RHIC data at 200 AGeV are found to be surprisingly consistent with
rather different $dE/dx$ models when coupled to recent 2+1D minimally viscous 
QGP flow field predictions. However, extrapolations to LHC, with parameters 
fixed at RHIC, favor running coupling QCD based energy-loss models 
over fixed coupling QCD, conformal AdS holography, or $T_c$-dominated 
jet-energy loss models that tend to overpredict jet quenching at the LHC.

\end{abstract}
\maketitle

{\em Introduction:} Jet quenching observables in high-energy 
nuclear collisions \cite{glv,JET} provide tomographic information 
about the density evolution of quark-gluon plasmas (QGP) but depend 
on details of jet-medium dynamics, $dE/dx(E,\vec{x},T)$, as well as 
on the bulk QGP collective temperature and fluid velocity fields, 
$[T(\vec{x},\tau),\vec{u}(\vec{x},\tau)]$. Here, $E$ is the energy of a jet
moving perpendicular to the beam axis at a transverse coordinate $x$ 
where the local temperature of the QGP is  $T$. In this Letter, we present 
predictions of a wide variety of models and compare to recent data
\cite{Adare:2012wg,Abelev:2012hxa,Chatrchyan:2012xq,CMS:2012aa,ATLAS:2011ah} 
on the nuclear modification factor $R_{AA}(p_T,\phi, 
\sqrt{s},b) \equiv dN_{AA}(b)/[N_{coll}(b) dN_{pp}]$ 
from the Relativistic Heavy Ion Collider (RHIC) and the Large Hadron Collider 
(LHC). We focus on the transverse momentum ($p_T$) and azimuthal angle ($\phi$) 
dependence of the high-$p_T>10$~GeV nuclear modification factor $R_{AA}$
where $N_{coll}(b)=T_{AA}(b)\sigma_{pp}^{in}$ is the average Glauber 
binary NN collision number for centrality classes $b$, corresponding 
to 0-5\% and 20-30\% at $\sqrt{s}=0.2\;{\rm and}\; 2.76$ ATeV.
We compare predictions of models based on perturbative QCD (pQCD), conformal AdS 
holography, and phenomenological $T\sim T_c\approx 170$~MeV dominated 
(SLTc) energy loss $dE/dx$ models coupled to different bulk QGP
temperature and collective velocity field evolution, $[T(\vec{x},t),
\vec{u}(\vec{x},t)]$, that include transverse and Bjorken longitudinal expansion.

\begin{figure*}[tbh]
\begin{minipage}[t]{8.5cm}
\hspace*{-1cm}
\includegraphics[width=3.5in]{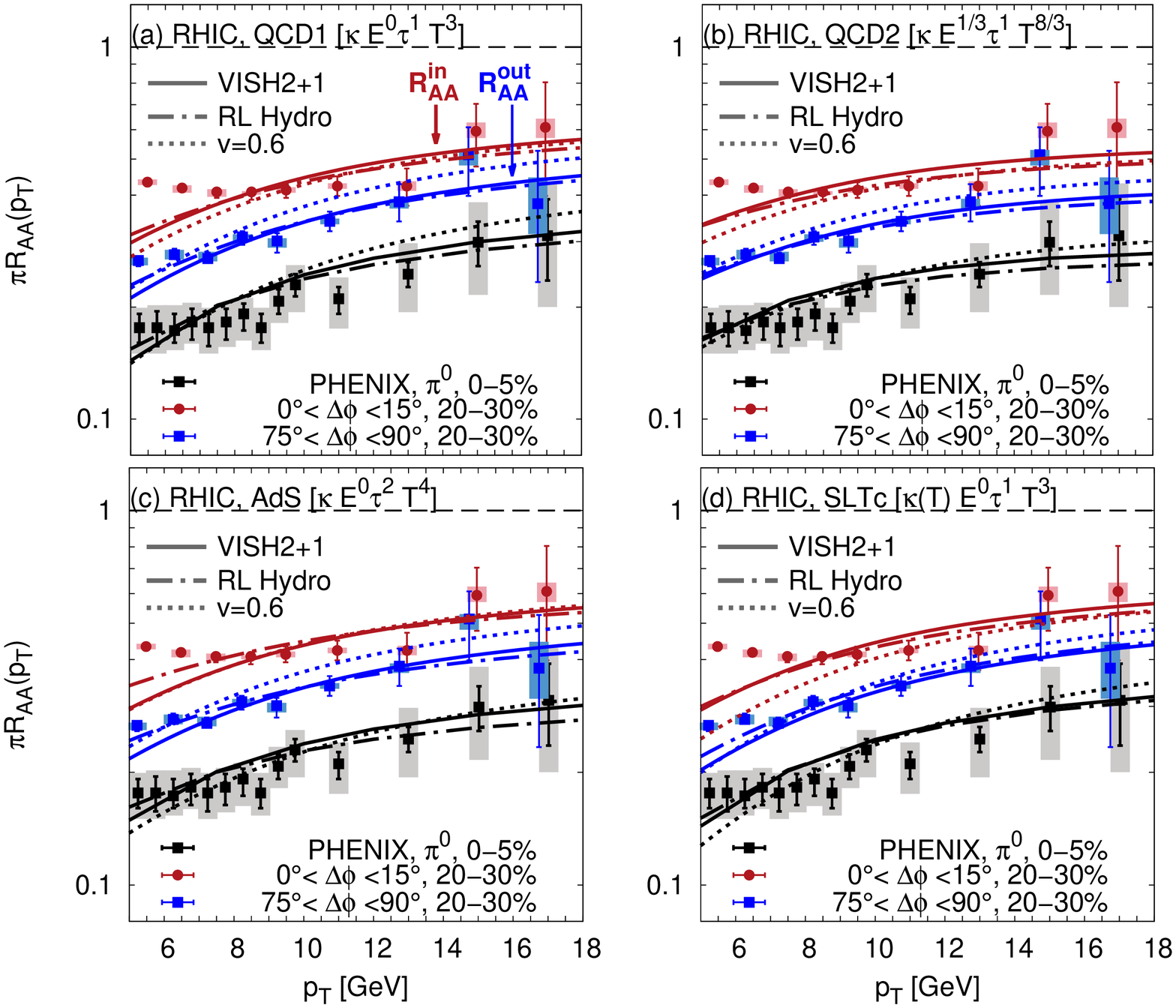}
\caption{Azimuthal jet tomography at RHIC. Panels (a-d) show PHENIX 200AGeV 
Au+Au data \protect{\cite{Adare:2012wg}} on $\pi^0$ 
nuclear modification factors in- and out-of-plane for 0-5\% and 20-30\% 
centralities, compared to predictions based on $dE/dx=\kappa(T) E^a x^b T^c$
\protect{\cite{WHDG11, Betz:2012qq}} for (a) QCD1 exponents (0,1,3) simulating a
QCD running coupling as in CUJET \protect{\cite{Buzzatti:2012dy}} and Refs.\ 
\protect{\cite{Betz:2012qq, Zakharov:2012fp}}, (b) QCD2 (1/3,1 8/3) describing 
a logarithmic jet-energy dependence as in fixed QCD coupling DGLV \protect{\cite{WHDG11,DGLV}}, 
(c) AdS (0,2,4) characterizing a conformal falling string energy loss as in Refs.\
\protect{\cite{Gubser:2008as,Marquet:2009eq}}, and (d) SLTc (0,1,3) with 
$\kappa(T_c) =3 \kappa(\infty)$ simulating a $T_c$-dominated energy loss 
as in Ref.\ \protect{\cite{Liao:2008dk}}. For each model, the quenching pattern 
is computed for three different bulk QGP fluid fields taken from:  
(1) ideal VISH2+1 \protect{\cite{Shen:2010uy}} (solid), 
(2) $\eta/s=0.08$ RL hydro \protect{\cite{Luzum:2008cw}} (dash-dotted), 
and (3) a $v_{\perp}=0.6$ blast wave model \protect{\cite{GVWH}} (dotted). 
In each case, the jet-medium coupling $\kappa$ is constrained by a fit to one 
single reference point $R_{Au+Au}^\pi(p_T=7.5\;{\rm GeV})=0.2$ of central $0-5\%$ 
Au+Au collisions.}
\label{Fig1}
\end{minipage}
\hspace*{0.7cm}
\begin{minipage}[t]{8.5cm}
\hspace*{-0.5cm}
\includegraphics[width=3.5in]{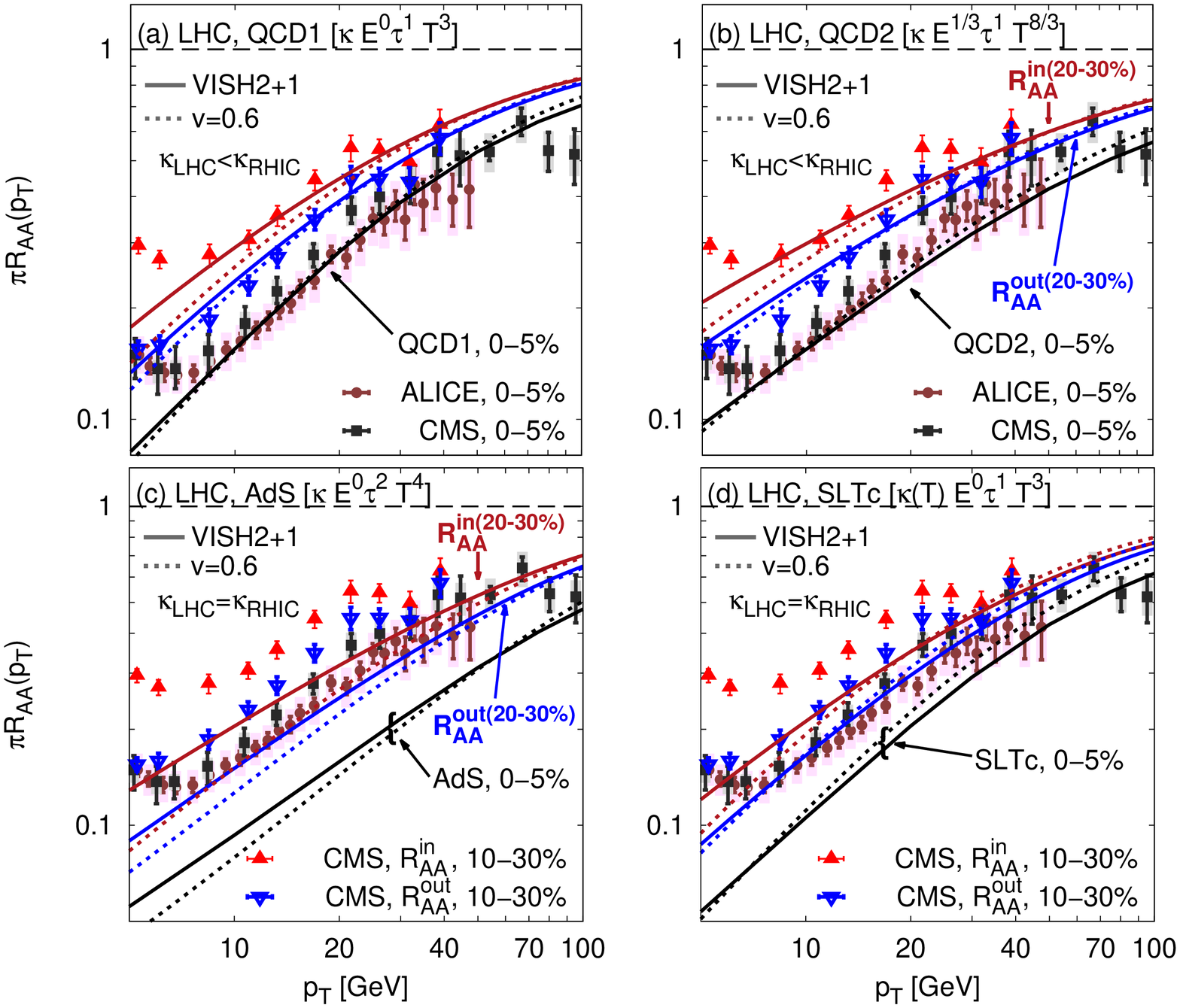}
\caption{Azimuthal jet tomography at the LHC 
\cite{Abelev:2012hxa,Chatrchyan:2012xq,CMS:2012aa,ATLAS:2011ah}. 
Panels (a-d) show ALICE \protect{\cite{Abelev:2012hxa}} (brown dots) and CMS 
\protect{\cite{Chatrchyan:2012xq}} (black squares) data on Pb+Pb collisions 
at 2.76 ATeV, compared to the same four $dE/dx$ models as Fig.\ \ref{Fig1} 
but using bulk QGP flow fields at LHC energies from viscous $\eta/s=0.08$
VISH2+1 \protect{\cite{Shen:2010uy}} (solid) and the $v_\perp=0.6$ blast wave
model \protect{\cite{GVWH}} (dotted). The jet-medium coupling $\kappa_{LHC}$ 
in (a) QCD1 and (b) QCD2 are reduced relative to RHIC to simulate 
running QCD coupling as in CUJET \protect{\cite{Buzzatti:2012dy}}.
In contrast, for {\em conformal} AdS \protect{\cite{Marquet:2009eq,Gubser:2008as}} 
in part (c) and a $T_c$-dominated SLTc model \protect{\cite{Liao:2008dk}} in part (d),
the {\em same} $\kappa$ is taken at LHC as fixed at RHIC.}
\label{Fig2}
\end{minipage}
\end{figure*}

The present work is motivated by recent PHENIX data \cite{Adare:2012wg} 
and the tentative conclusions drawn that AdS/CFT motivated jet-energy loss 
$dE/dx=\kappa x^2T^4$ models \cite{Bass:2008rv,Marquet:2009eq,Jia:2011pi} with 
particular assumptions about the QGP $(T,\vec{u})$-fields seem to describe the 
latest RHIC data better than QCD-based models. The RHIC data shown in Fig.\ 
\ref{Fig1} are in- and out-of-plane nuclear modification factors, 
$R_{AA}^{in} \equiv R_{AA}(0<\phi<15^\circ )$ and 
$R_{AA}^{out} \equiv R_{AA}(75^\circ <\phi<90^\circ)$. Black squares are 
0-5\% data and red (blue) symbols are $R^{in}_{AA}(R^{out}_{AA})$ data at 
20-30\% centrality. The aim of the present Letter is to  test the robustness of 
the PHENIX conclusion by considering a wider class of $dE/dx$ models coupled to 
different QGP flow fields as well as to extend the analysis to a simultaneous 
description of both RHIC and LHC data. With an order of magnitude higher 
$\sqrt{s}$, the LHC can probe much higher $p_T$ ranges as well as more than doubled
QGP densities $\propto T^3$ as compared to RHIC. In addition, the initial invariant jet-production 
distributions at $y=0$, $g_r(p_T)=dN_r^{jet}/dyd^2p_T$, for $r=q,g$ jets 
changes by orders of magnitude from RHIC to LHC. Therefore, cross comparison 
of RHIC and LHC data provides the most stringent  tests so far 
of the consistency and quantitative power of proposed models of jet-energy loss and 
space-time density evolution of the bulk QGP produced in ultrarelativistic 
nuclear collisions.
 
Both, magnitude and azimuthal dependence of jet quenching in non-central 
collisions are most conveniently 
studied via $R_{AA}^{in/out}$ \cite{Bass:2008rv}. These 
observables are sensitive to details of the jet energy, path length, and 
temperature dependence of $dE/dx$ (see, e.g.\ Refs.\
\cite{GVWH, Renk:2011aa,Renk:2011ia, Renk:2011gj,Chen:2011vt, Molnar:2013eqa}).
In particular, they depend on the details of the QGP transverse expansion
\cite{Song:2007ux,Shen:2010uy,Luzum:2008cw, Niemi:2008ta,Shen:2011eg,Qiu:2011hf},
as especially emphasized in Refs.\ \cite{Renk:2011aa,Molnar:2013eqa}. We constrain 
each model by fitting the jet-medium coupling $\kappa$ to a single reference 
point at $p_T=7.5$~GeV for central 0-5\% Au+Au at $\sqrt{s}=200$~AGeV with 
$R_{AA}=0.20$, as in Ref.\ \cite{Bass:2008rv}. We extend our previous work 
\cite{WHDG11, Betz:2012qq, Buzzatti:2012dy} by taking into account a 2+1D 
transverse expansion as predicted by (1) VISH2+1 
\cite{Song:2007ux, Shen:2010uy, Shen:2011eg, Qiu:2011hf}, (2) RL Hydro 
\cite{Luzum:2008cw}, 
and (3) a simple $v_{\perp}=0.6$ transverse blast wave model \cite{GVWH} for reference.
We further broaden the PHENIX analysis \cite{Adare:2012wg} by considering 
also the non-perturbative model of energy loss SLTc \cite{Liao:2008dk} 
that postulates the dominance of the transition temperature region 
$T\sim T_c\approx 170$ MeV.

In order to interpolate between QCD, AdS/CFT, and $T_c$-enhanced models
of energy loss, we utilize a convenient parametric model 
\cite{WHDG11,Betz:2012qq} of $dE/dx$ characterized by three exponents $(a,b,c)$
that control the jet energy, path length, and thermal-field dependence, and allow 
for the possibly that the jet-medium coupling, $\kappa(T)$, could depend 
non-monotonically on the local temperature field as in the SLTc model:
\begin{eqnarray}
\hspace*{-3ex}
\frac{dE}{dx}=\frac{dP}{d\tau}(\vec{x}_0,\phi,\tau)= 
-C_r\kappa(T)  P^a(\tau) \, \tau^{b} \, T^c
\;,
\label{Eq1}
\end{eqnarray}
where $T=T[\vec{x}(\tau)=\vec{x}_0+ (\tau-\tau_0) \hat{n}(\phi),\tau]$ is the local 
temperature along the jet path at time $\tau$ for a jet produced 
initially at time $\tau_0$ and distributed according to either a 
Glauber or a KLN transverse initial profile. In Eq.\ (\ref{Eq1}), 
$C_r=1(\frac{9}{4})$ describes quark (gluon) jets. For jets of type $r=q,g$ 
produced with invariant transverse momentum distribution, $g_r(P_0)$, taken 
from Refs.\ \cite{WHDG11,Betz:2012qq}, the nuclear modification factor is given by 
$R_{AA}^r(P_f,\phi)=g_r(P_0(P_f,\phi)]/g_r(P_f)(dP^2_0/dP^2_f)$. The initial 
jet energy (prior to fragmentation), $P_0$, is then related to the final 
quenched energy, $P_f$, by
\begin{eqnarray} 
\hspace*{-1ex} 
P_0(P_f,\phi)=\left[P_f^{1-a}+ \int_{\tau_0}^{\tau_f}
K(T) \tau^b T^{c}[\vec{x}_\perp(\tau),\tau]d\tau
\right]^\frac{1}{1-a}\hspace*{-2ex},
\label{Eq2}
\end{eqnarray} 
where the effective coupling is $K(T)=(1-a)C_r \kappa(T)$. Eq.\ (\ref{Eq2}) 
illustrates the competition between the intrinsic $dE/dx\propto E^ax^bT^c$ 
and the local hydrodynamic temperature field dependence including a possible 
non-monotonic jet-medium coupling $\kappa(T)$.

{\em RHIC and LHC Results:} Perturbative QCD based models labeled QCD1 and QCD2 
in Figs.\ 1a, 1b and 2a, 2b correspond to exponents of $(0,1,3)$ and $(1/3,1,8/3)$. 
QCD1 with $a=0$ simulates the effects of a running QCD coupling as found with CUJET 
\cite{Buzzatti:2012dy}. QCD2 assumes $a=1/3$ to simulate a logarithmic energy 
dependence predicted with fixed QCD coupling 
\cite{glv,DGLV, WHDG11,Betz:2012qq,Buzzatti:2012dy}), but also allows a
 $\kappa_{LHC}<\kappa_{RHIC}$.  In Refs.\ \cite{Betz:2012qq,Buzzatti:2012dy}, 
the opacity integral in Eq.\ (\ref{Eq2}) was evaluated taking only 1+1D Bjorken 
expansion with $v_\perp=0$ into account. In Figs.\ 1 and 2 the opacity integrals 
were evaluated with three variants of transverse flow fields: (1) ideal VISH2+1 
\cite{Song:2007ux}, (2) viscous RL hydro \cite{Luzum:2008cw}, and (3) a 
$v_\perp=0.6$ blast wave flow \cite{GVWH} using a radial 
$r(t)=(1+v_\perp^2\tau^2/R^2)^{1/2}$ dilation of the initial transverse profile:
$\rho(x,y,\tau)=\rho_0[x/r(\tau),y/r(\tau)][\tau_0/\tau r^2(\tau)]$.
Here, $R$ denotes the mean radius. 
As noted above, for each model of transverse flow, the jet-medium coupling at 
RHIC was adjusted to fit a single reference point, as in Ref.\ \cite{Bass:2008rv}.

The most striking result in Fig.\ 1a is that in contrast to the (AMY,
HT, and ASW) pQCD models \cite{Bass:2008rv}, compared to data of
Ref.\ \cite{Adare:2012wg}, the QCD1 model combined with either ideal
VISH2+1 or viscous RL hydro transverse flow agree within present
errors with RHIC data in the high-$p_T> 8$ GeV region.  However, QCD1
in a $v_\perp=0.6$ transverse blast wave background leads, as in Ref.\
\cite{GVWH} with $v_\perp =0$, to an in/out asymmetry with a factor of two
below recent PHENIX data \cite{Adare:2012wg}.
Ref.\ \cite{Molnar:2013eqa} also found that a
GLV $dE/dx$ \cite{glv} evaluated in the MPC parton cascade background
underpredicts the high-$p_T$ elliptic asymmetry observed at RHIC
that was another major motivation for the present work.

The difference between models shown in Fig.\ 1a and Fig.\ 22 (a,b,c)
of Ref.\ \cite{Adare:2012wg} is due to different combined effects of
$dE/dx$ and bulk QGP flow. In Ref.\ \cite{Bass:2008rv}, the flow field
was computed with an ideal (non-dissipative) hydrodynamic code
assuming a Bag model first order-phase transition with vanishing speed
of sound over a wide energy density range. Here however, the VISH2+1 grid used
in Fig.\ 1(a-d), utilizes a smoothed (SM-EOS Q) equation of state
(EoS), while the viscous RL hydro employs a more realistic continuous
crossover transition EoS. We checked (not shown) that minimal viscous
VISH2+1 temperature fields lead to less than $10\%$ variations from
the ideal VISH2+1 hydro predictions shown in Fig.\ \ref{Fig1}.
 
However, we cannot interpret the approximate agreement of QCD1 with
RHIC data as success because in fact we found that all four $dE/dx$
models in (a-d) performed equally well at RHIC when coupled with
VISH2+1 and RL backgrounds. Please note in fact that all four models perform
equally poorly in the reference $v_{\perp}=0.6$ blast wave background.

The difficulty of untangling the dE/dx and QGP flow field effects 
 at one particular $\sqrt{s}$ leads us to consider the 
higher discriminating power afforded by exploiting
the dependence of $R_{AA}$ on $\sqrt{s}$ in the range $0.2-2.76$ ATeV.
In the case of QCD1, we find in Fig.\ 2a that 
the predictions agree within present errors at LHC both in magnitude 
and $p_T$-slope of $R_{AA}$ when a $(a=0, b=1, c=3)$ loss is coupled to viscous 
VISH2+1 LHC fields. As in Fig.\ 1, the reference $v_\perp=0.6$ blast wave flow leads 
to a significant underestimate of the azimuthal asymmetry at LHC energies
as also predicted with GLV \cite{glv} coupled to MPC parton transport theory 
in Ref.\ \cite{Molnar:2013eqa}.

In Fig.\ 1b we found that at RHIC there is very weak
sensitivity to the jet $E^a$-dependence in the range $a=0-1/3$, but at LHC
the larger $p_T$ slope of $R_{AA}$ favors QCD1 ($a=0$) over QCD2 ($a=1/3$), 
and supports the running coupling explanation proposed 
with CUJET \cite{Buzzatti:2012dy} albeit in  $v_\perp=0$ backgrounds.  
It is important to note that in {\em both} QCD1 and QCD2 cases the 
jet-medium coupling $\kappa$ has been reduced by $\sim 30\%$ from 
their constrained values at RHIC. This reduction is natural in 
perturbative QCD based $dE/dx$ due to running of the combined radiation and
scattering couplings, 
$\kappa_{QCD} \propto \alpha_s\{k_\perp^2/[x(1-x)]\}\alpha_s^2(q^2)$, in the
DGLV opacity series integrals \cite{DGLV} over the
radiated gluon momentum fraction $x$, the gluon transverse momentum $k_\perp$, 
and the medium momentum transfers $q$ generalized 
in CUJET \cite{Buzzatti:2012dy} to include running QCD coupling effects.
See Ref.\ \cite{Zakharov:2012fp} for the path integral formulation of this 
problem. 
The free parameter set in CUJET to fit the RHIC reference point is the maximum
$\alpha_s^{max}=\alpha_s(Q^2<1\;{\rm GeV}^2)=0.4$.

In contrast to the consistent account of both RHIC and LHC data
by QCD1 combined with viscous VISH2+1 flow in Figs.\ 1a and 2a, 
the {\em conformal} AdS-inspired model \cite{Marquet:2009eq} for 
$dE/dx\equiv \kappa x^2T^4$ in the same background fails the
extrapolation from RHIC to LHC with fixed $\kappa$ fit to RHIC data. 
In true AdS/CFT, $\kappa\propto \sqrt{\lambda}$, where  
$\lambda = 4\pi \alpha_s N_c$ is the 'Hooft coupling,
and the applicability of classical gravity holography requires
$\lambda\gg 1$.  However, in {\em conformal} AdS/CFT, $\lambda$
cannot run. Once fixed at RHIC, the AdS falling string model \cite{Gubser:2008as}
overpredicts LHC quenching as shown in Ref.\ \cite{Ficnar:2012yu},
even for $\lambda$ as low as $1$ in static backgrounds and
even if quadratic curvature corrections are taken into
account. To fit both RHIC and LHC data in this AdS scenario, 
$\kappa$ needs to be reduced by a factor of two from RHIC to LHC
\cite{Betz:2012qq} which is inconsistent with assumed conformal invariance.
We conclude that consistency between RHIC and LHC jet tomography
will require  at least further generalization of present 
holographic jet quenching models to allow for more general 
string initial conditions and  non-conformal geometric 
deformations \cite{ficnar13,Mia:2012yq}.

Finally, we consider the class of $dE/dx$ models, labeled SLTc
\cite{Liao:2008dk}, that assume the dominance of energy loss in
regions of the QGP with $T\sim T_c$.  One such model is based on a
scenario that associates the QCD conformal anomaly near $T_c$ with
color magnetic monopole condensation. Scattering of color electric
charged jets by color magnetic monopoles could lead to an enhancement
of $dE/dx$ in the QCD crossover transition regions that have higher
spatial elliptic eccentricity than the average. We simulate this
effect in Figs.\ 1d and 2d by using the simplest step function model of Ref.\
\cite{Liao:2008dk} for the local jet-medium coupling with $\kappa(113
<T<173\;{\rm MeV}) = \kappa_c =3\kappa_Q $ and
$\kappa_Q=\kappa(T>173)$. For $\kappa_c/\kappa_Q=3$, the fitted value
of $\kappa_Q$ to the RHIC reference point leads to the same
satisfactory description of the 0-5\% as well the 20-30\% RHIC data
for $p_T>8$ GeV as the other models in parts (a-c). Note that our
SLTc calculations generalize those of Ref.\ \cite{Liao:2008dk} by coupling the 
model to the three different transverse flow fields shown in
Fig.\ 1d and by testing both the $p_T$ and $\phi$ dependence of $R_{AA}$. 
 
When extrapolated to LHC  with fixed $\kappa_c = 3 \kappa_Q$, we find 
in Fig.\ 2d the same problem with SLTc as with an AdS-like model in Fig.\ 2c, 
namely, an overprediction of the magnitude jet quenching at all centralities.
We have not attempted more general $\kappa(T,\sqrt{s})$ variations of the SLTc 
models since we found in Fig.\ 1a and 2a that QCD1 with 
$\kappa_c=\kappa_Q$, corresponding most closely to running coupling QCD 
\cite{Buzzatti:2012dy,Zakharov:2012fp}, adequately accounts for both RHIC 
and LHC data within present errorbars considering one $\alpha_{max}$ parameter.

{\em Conclusions:} We compared recent data on the nuclear modification factor
measured at RHIC \cite{Adare:2012wg} and LHC energies 
\cite{Abelev:2012hxa,Chatrchyan:2012xq} to a wide class of jet-energy loss
models describing (a) a pQCD-like energy loss with running coupling 
\cite{Buzzatti:2012dy}, (b) a QCD-like, similar logarithmic energy loss \cite{WHDG11},
(c) an AdS/CFT-inspired energy loss, and (d) a $T_c$-dominated energy-loss
model (SLTc) \cite{Liao:2008dk} in different transverse expanding, collective 
flow backgrounds. Comparing RHIC and LHC results, we found that for a 
realistic, transverse expanding medium, running coupling perturbative 
QCD energy loss seems to be favored. We note however that at both, RHIC and LHC, 
the magnitude of $R_{AA}$ in the intermediate (``IM'') $2<p_T<8$~GeV kinematic 
region is underpredicted by all jet-quenching models considered here. This ``IM''
region interpolates between the perfect fluid low-$p_T<2$~GeV infrared (``IR'')
range and the high-$p_T>8$~GeV ultraviolet (``UV'') perturbative QCD quenched 
jet range. A proper theory of  jet quenching in the non-equilibrium QGP ``IM'' 
range remains a formidable challenge. Further details of the present study 
will be presented elsewhere.

{\em Acknowledgments:} We are especially grateful to P.\ Romatschke, U.\ Heinz, 
and C.\ Shen for making their hydrodynamic field grids available.
Discussions with A.\ Buzzatti, A.\ Ficnar, J.\ Harris, W.\ Horowitz, 
J.\ Jia, J.\ Liao, M. Mia, D.\ Molnar,
G.\ Torrieri,  and X.-N.\ Wang have been particularly valuable.
BB acknowledges financial support received
from the Helmholtz International Centre for FAIR within the framework
of the LOEWE program (Landesoffensive zur Entwicklung
Wissenschaftlich-\"Okonomischer Exzellenz) launched by the State of
Hesse. This work was supported in part from the US-DOE Nuclear Science Grant 
No.\ DE-FG02-93ER40764 and No.\ DE-AC02-05CH11231 within the framework of the 
JET Topical Collaboration \cite{JET}.

\end{document}